\documentstyle[stwol,epsf]{article}
\input{psfig}
\def\Journal#1#2#3#4{{#1} {\bf #2}, #3 (#4)}

\def\NPB{{\em Nucl. Phys.} B}
\def\PLB{{\em Phys. Lett.}  B}
\def\PRL{{\em Phys. Rev. Lett.}}
\def\PRD{{\em Phys. Rev.} D}
\def\ZPC{{\em Z. Phys.} C}

\def\be{\begin{equation}}
\def\ee{\end{equation}}
\def\bea{\begin{eqnarray}}
\def\eea{\end{eqnarray}}

\def\gevc{GeV$/c$}
\begin{document}
\title{QUARKONIUM PRODUCTION IN $p\overline{p}$
COLLISIONS AT THE TEVATRON}
\author{ SLAWEK M. TKACZYK\footnotemark%
}
\address{Fermi National Accelerator Laboratory, P. O. Box 500,
Batavia, IL 60510}
\twocolumn[\maketitle\abstracts{
 Charmonium and bottomonium production is studied using $\mu^+\mu^-$ data
samples collected by the CDF and D0 experiments during the 1992-96 $p\overline{p}$ 
collider run at the Fermilab Tevatron. 
  The inclusive cross sections as 
a function of the transverse momentum of reconstructed quarkonium 
states are measured. The results are compared with theoretical predictions,
 which take into account different quarkonium production mechanisms.}]

\footnotetext{Representing the CDF and D0 Collaborations.}%

\section{Introduction}

During the 1992-96 $p\overline{p}$
collider Run I, the Tevatron delivered over 150 pb$^{-1}$ of integrated luminosity.
The charmonium and bottomonium production rate results presented in this paper
were measured by the CDF and D0 collaborations and are based on dimuon subsets of the  
data.
 In the
quarkonium analyses, differential cross sections for $J/\psi$, $\psi(2S)$ and
three $\Upsilon$ states have been measured using their $\mu^+\mu^-$ decay
channels. Two muons in the final state were used as a trigger signature.

In early studies of quarkonium production the dominant contributions were
assumed to come from the lowest order Feynman diagrams involving gluon fusion 
---
either directly into quarkonium states and recoiling partons or through a
$b\overline{b}$ pair followed by decays $B\rightarrow  \Psi X$ --- or,
 in the case
of the $J/\psi$ or the $\Upsilon$,  radiative decays of $\chi_c$ or $\chi_b$ 
mesons.\footnote{$\Psi$ represents both $J/\psi$ and $\psi(2S)$ mesons.}
Various methods are used
to disentangle the three sources,  providing information  about charm and
bottom production and fragmentation mechanisms at low transverse momenta.

Previous CDF measurements of $J/\psi$ and $\psi(2S)$ production rates during the
1988-89 collider {run\cite{cdf89}} showed production cross sections
considerably larger than contemporary theory predicted. This drew theoretical
interest, but at that time the question of whether or not the excess could be
attributed to a large prompt component was not addressed. It has been
pointed out that in addition to gluon fusion, the gluon fragmentation processes
are also important sources of quarkonium { production.\cite{braatrev}} We
briefly summarize the  proposed improvements in the theoretical description of
the observed production rates.  

\section{Charmonium total cross sections}
Muons identified in the central pseudorapidity range, 
 $|\eta|<0.6$, were used in these analyses. Additionally, 
 the D0 detector provides muon 
coverage in the forward pseudorapidity region, allowing production rate
studies in the range  $2.5<|\eta|<3.7$.

In the CDF analyses, both muons of the reconstructed $\Psi$ candidate are
required to have $p_T$ greater than 2.0 GeV$/c$, and at least one muon of the
pair is required to have $p_T > 2.8$ GeV$/c$. The reconstructed $J/\psi$ or
$\psi(2S)$ candidates must have $p_T > 5.0$ GeV$/c$. About 22,000 $J/\psi$ events
and 800 $\psi(2S)$ events are reconstructed in data samples of 15.4 pb$^{-1}$
and 17.8 pb$^{-1}$, respectively.  In the D0 analysis, muons with transverse
momentum greater than 3 \gevc are reconstructed in the central and forward 
pseudorapidity regions, and dimuon candidates are required  to have 
$p_T> 8$ GeV/$c$. 
About 4,000 $\Psi$ events in the central and about 500 events in the
forward rapidity ranges are found from a fit to a Gaussian function and 
physics-motivated background in the samples of 60 pb$^{-1}$ and 9.3 pb$^{-1}$, 
respectively. 
The D0 tracking momentum resolution is such that it does
not allow separation of the $\psi(2S)$ and $J/\psi$ states.

 CDF measures the product of dimuon branching ratio times integrated cross 
section to be 
\mbox {$17.35\pm0.14\pm2.79$ nb} for
$J/\psi$ and \mbox{$0.57\pm0.04\pm0.09$ nb} for $\psi(2S)$ in the central 
rapidity 
region with $p_T > 5.0$ \gevc,  
and D0 measures
\mbox {$1.96\pm0.16\pm0.63$ nb} for $J/\psi$ in the central region
 and \mbox{$0.40\pm0.04\pm0.04$ nb} in the forward 
 region with $p_T > 8.0$ \gevc.

\section{Charmonium from $b$ Decay}

\begin{figure}[h]
\center
\psfig{figure=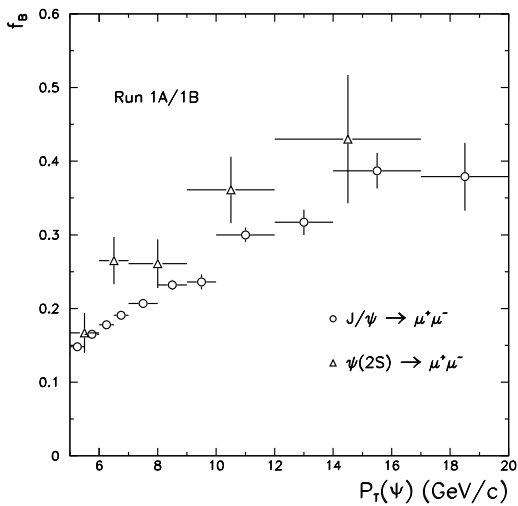,height=2.75in} 
\caption{Fraction of $J/\psi$ and $\psi(2S)$ from $b$ decay as a function 
of $p_T$.
\label{myfig1}}
\end{figure}

\begin{figure}[h] 
\leftline{
\epsfysize 6.5cm
\epsffile{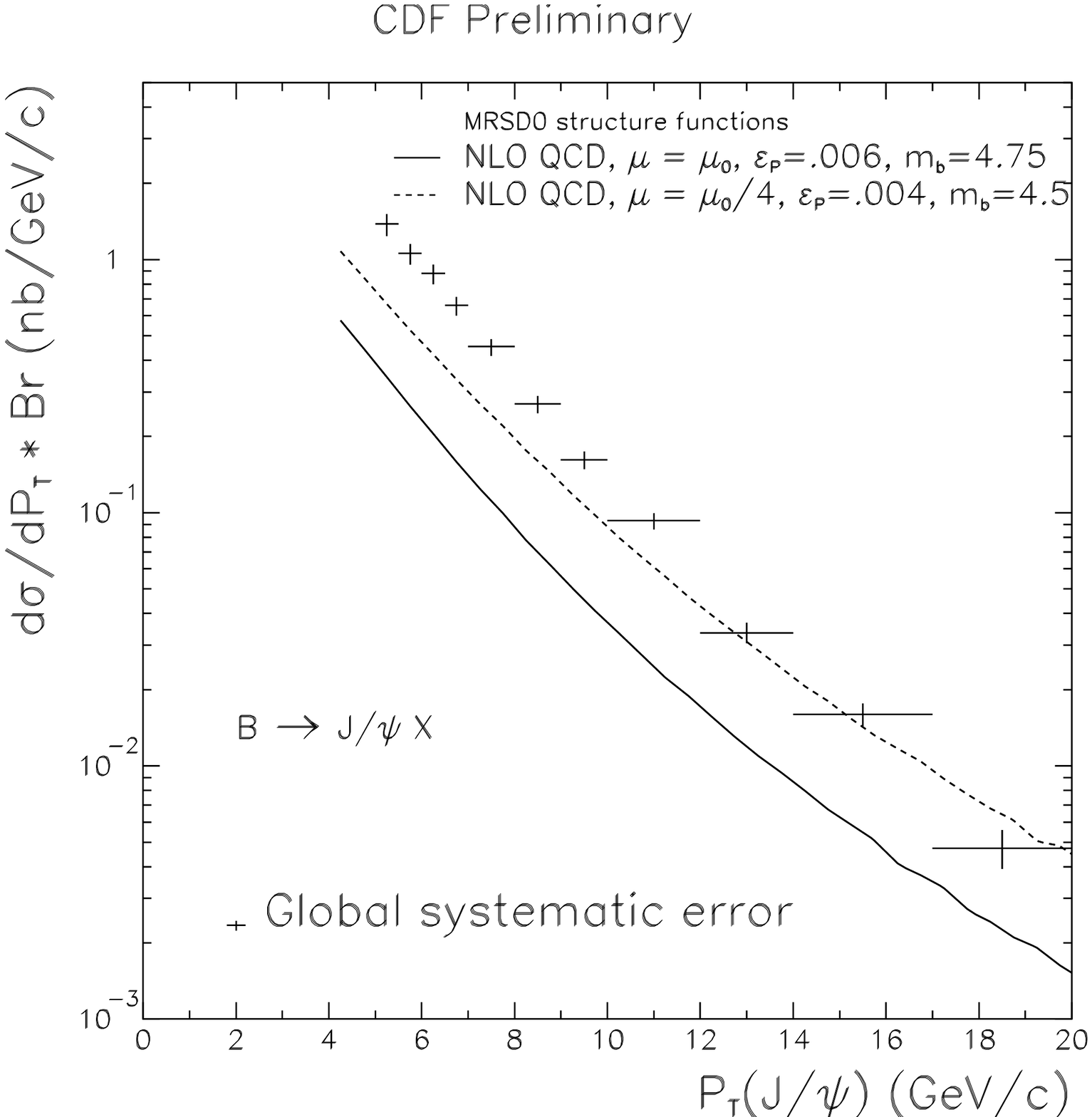}}
\leftline{
\epsfysize 6.5cm 
\epsffile{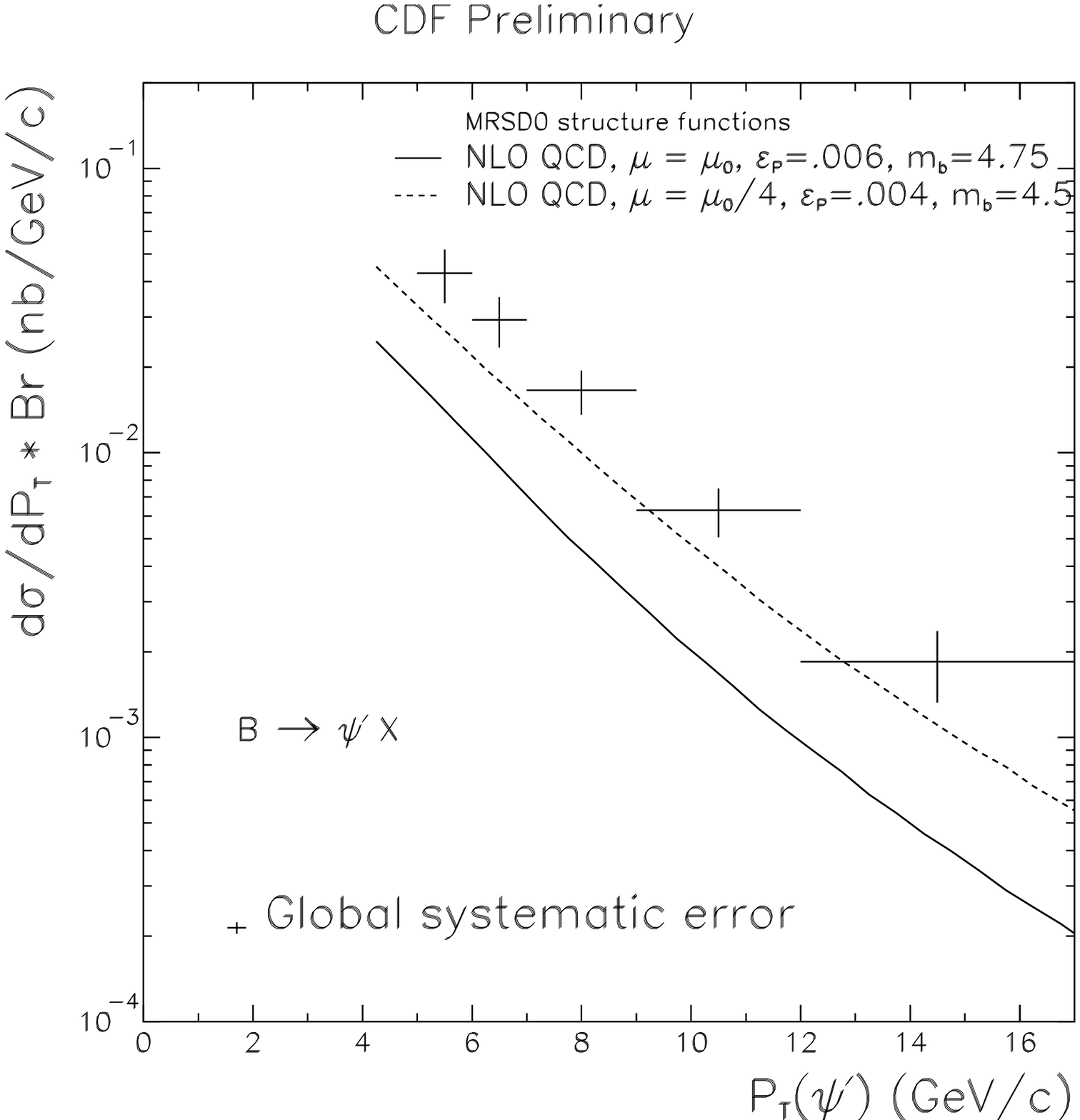} }
\caption{$b$ quark cross section determined from the $J/\psi$ (top) and 
$\psi(2S)$ (bottom) samples. 
\label{myfig3}} 
\end{figure}

The CDF collaboration, using a silicon vertex detector,  separated the $J/\psi$
and $\psi(2S)$ samples into their components arising from $b$ decay and from
prompt production by analyzing the proper decay-length ($c\tau$) distributions.
The $c\tau$ distribution is fitted to three components:  an exponential
convoluted with a Gaussian resolution function for the $b$ hadron decay
component, a Gaussian function centered at 0 for prompt production, and a 
Gaussian function with positive and negative exponential tails to describe the
background, both combinatorial as well as from  sequential
$b\rightarrow\mu^- c \rightarrow \mu^+ s$ decays.  The samples are subdivided
into ranges of $p_T(\mu^+\mu^-)$ and fitted separately for each range.  Figure
\ref{myfig1} shows the $J/\psi$ and $\psi(2S)$ fractions from $b$ decay as
 a function of $p_T$.  
  These fractions are then convoluted
with the charmonium $p_T$ spectra to give the $b$ cross sections, as shown in
figure \ref{myfig3}.  The results are within a factor of 2-3
of the NLO QCD \mbox{prediction,\cite{nde}} using a central value of the 
input
 parameters,\footnote{The input parameters have the following values: renormalization and fragmentation scales
 $\mu_F=\mu_R=\mu_0=\sqrt(m_b^2+p_T^2)$, 
the mass of the $b$ quark,
 $m_b$=4.75 GeV/$c$, the Peterson fragmentation parameter, 
$\epsilon_b = 0.006$.}
 and are
consistent with other Tevatron $b$ cross section \mbox{results.\cite{myprl}}

\section{Prompt Charmonium Production}

The cross section for prompt charmonium production is in disagreement with
theoretical predictions based on color-singlet production of bound 
$c\overline{c}$
 \mbox{states.\cite{gms}}  The rate of prompt $\psi(2S)$ production is
about a factor of 50 larger than predictions based on such a model.  All prompt
$\psi(2S)$ are believed to be directly produced since $\chi_c$ states with
sufficient mass to decay to $\psi(2S)$ lie above the threshold for strong decays
to $D\overline{D}$ meson pairs.  However, prompt $J/\psi$ are produced not only
directly, but also via $\chi_c$ radiative decays. 
CDF has determined the fraction of the prompt $J/\psi$ sample coming from 
the $\chi_c$ decay by fully reconstructing the decay  
$\chi_c \rightarrow J/\psi \gamma$.
Photon candidates detected in the central electromagnetic calorimeter with
energy greater than 1 GeV and having no charged track pointing to the same
calorimeter tower are combined with $\mu^+\mu^-$ pairs consistent with the J/$\psi$ mass. 
 A peak containing $1230\pm72$ $\chi_c$ candidates is
observed in the mass difference
distribution $M(\mu^+\mu^-\gamma)-M(\mu^+\mu^-)$. 
The background under the peak has been modelled
by embedding simulated $\pi^0$ and $\eta^0$ decay photons in real $J/\psi$
events.  The fraction of $J/\psi$ coming from prompt $\chi_c$ decay,
measured for 4
different $p_T$ bins, ranges from about $32\%$ in the 4-6 GeV/$c$ bin to
$28\%$ in the bin $p_T>10$ GeV/$c$. The D0 measurement of this 
fraction is consistent with these \mbox{results.\cite{d0chi}}
Multiplying the total prompt  $J/\psi$ 
cross section by the $\chi_c$ fraction shows that the rate of $J/\psi$
production from $\chi_c$ is within a factor of 2-3 of the theoretical
prediction, but, as with the $\psi(2S)$, the remaining direct $J/\psi$ cross
section is about a factor of 50 larger than the color-singlet prediction.
\begin{figure}[h] 
\center
\psfig{figure=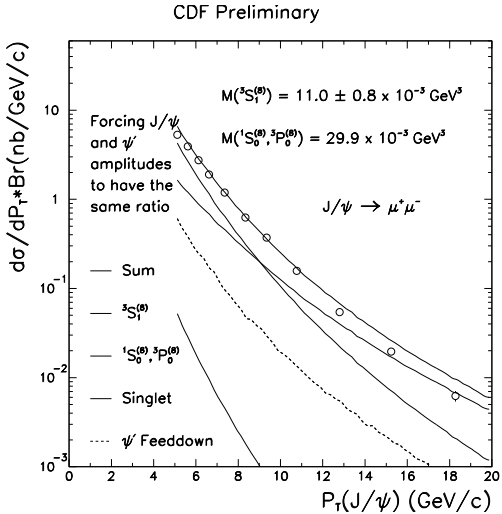,height=2.65in} 
\caption{Prompt direct $J/\psi$ sample, compared to the theoretical
prediction, with color-octet components fitted simultaneously to the
$J/\psi$ and $\psi(2S)$ distributions.
\label{myfig5}}
\end{figure}
\begin{figure}[h] 
\center
\psfig{figure=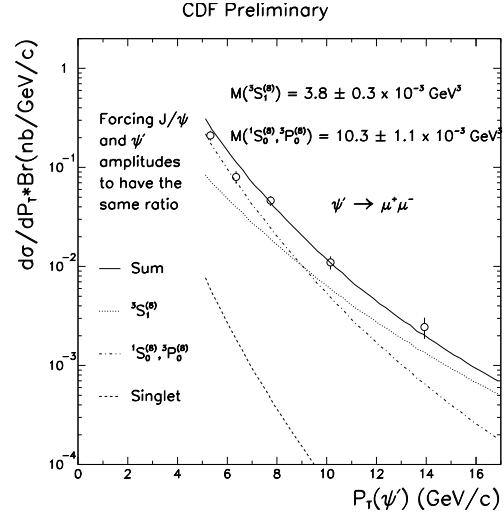,height=2.65in} 
\caption{Prompt direct $\psi(2S)$ sample, compared to the theoretical
prediction, with color-octet components fitted simultaneously to the
$J/\psi$ and $\psi(2S)$ distributions.
\label{myfig6}}
\end{figure}

     One proposal to explain the observed prompt charmonium production rates
 is to include $c\overline{c}$ pairs produced in a color-octet
\mbox{state.\cite{braflem}}  The initial
production can be calculated perturbatively and can be used to predict the
$p_T$ dependence of the cross section.  The transition to a color-singlet
state needed to form a bound $c\overline{c}$ particle proceeds via soft gluon
emission.  This latter process cannot be calculated perturbatively, so the
normalization is found by fitting the theory to the data.  Figures \ref{myfig5}
and \ref{myfig6}
show the prompt $J/\psi$ and $\psi(2S)$ cross sections and the
corresponding theoretical predictions when the fitted color-octet contributions
are \mbox{included.\cite{cho}} The predictions of the model can be further tested
by fixed target  hadro- and photo-production experiments and by measurement of 
 the $\psi(2S)$ polarization in $p\overline{p}$  collisions. 

\section{Forward Charmonium Production}

\begin{figure} 
\leftline{
\epsfysize 4.0cm
\epsffile{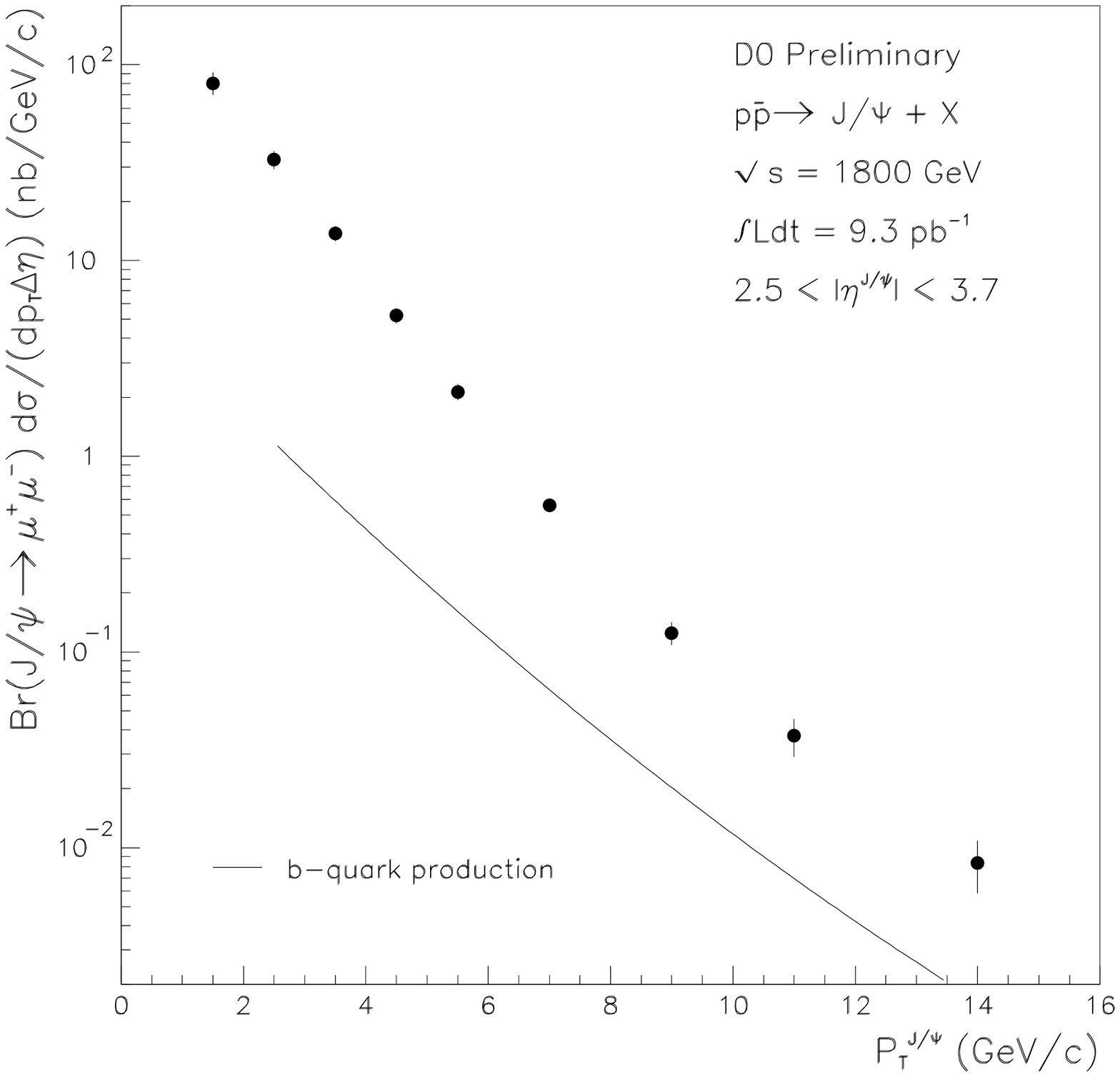}
\epsfysize 4.0cm 
\epsffile{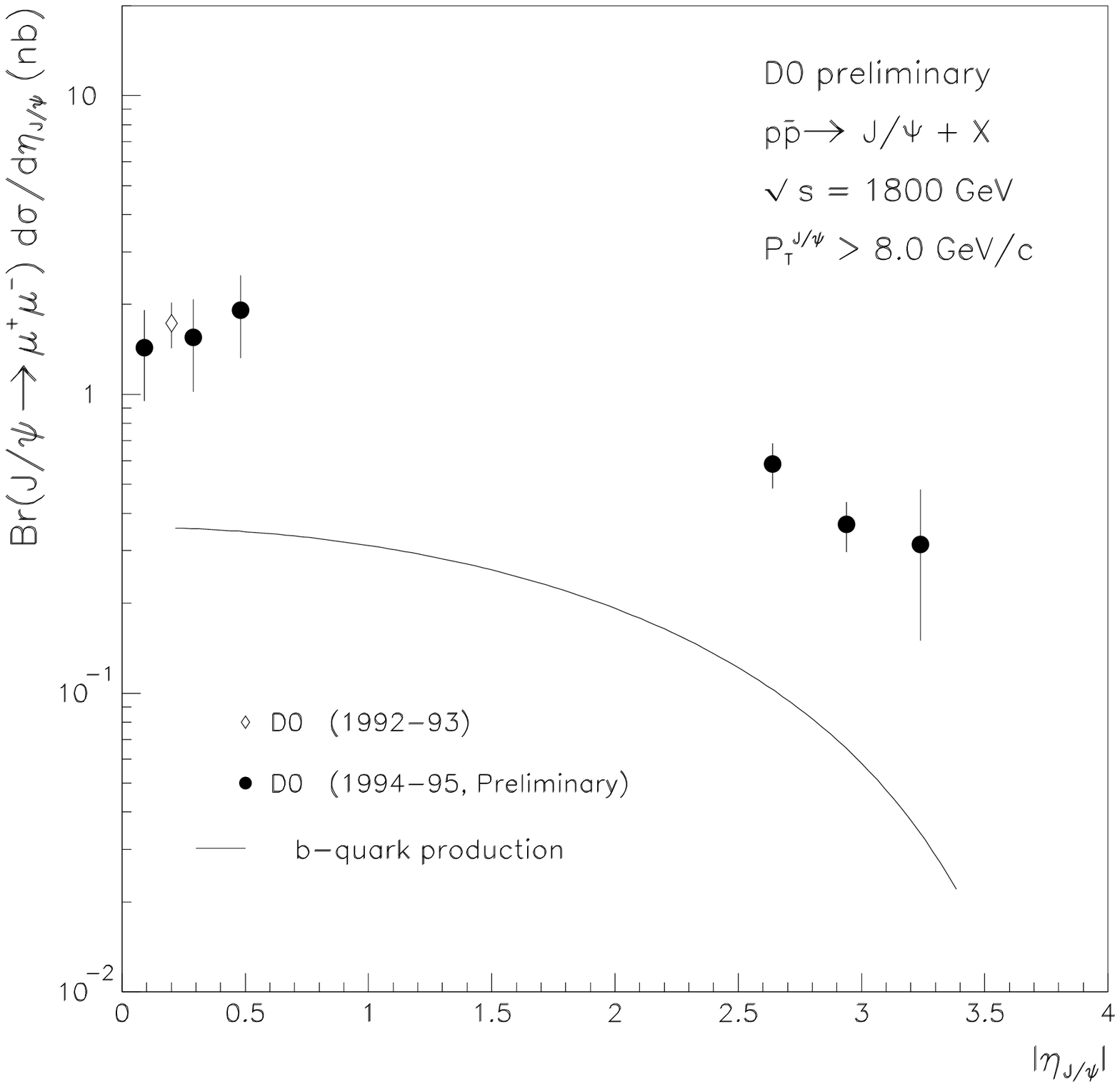} }
\caption{The $\Psi$ cross sections - Br$\cdot d\sigma/dp_T$ 
 for $2.5<|\eta|<3.7$ (left),
and Br$\cdot d\sigma/d\eta$ for $p_T>8$ \gevc (right).
The solid curve represents the expected contribution from
$b$ quark fragmentation.
\label{d0fig1}} 
\end{figure}

The D0 collaboration determined
the differential cross section, $d\sigma/dp_t$, shown in 
Figure \ref{d0fig1}, using  $\Psi$ candidates reconstructed in the forward 
\mbox{region.\cite{d0for}} This is the first measurement of the $\Psi$ cross
section at large pseudorapidity. Figure \ref{d0fig1} combines data from the 
central and forward $\Psi$ analyses for transverse momentum greater 
than 8 \gevc. The
 measurements in the central region are consistent with
having no $\eta$
dependence. However in the forward region the measured cross section 
is approximately a factor of 5 lower than the central region values.
 The data points 
 are compared in Figure  \ref{d0fig1} with a preliminary theoretical 
prediction of the contribution from the $b$ 
quark decay to $\Psi$.
 
\section{Bottomonium Production}

     The CDF collaboration has also 
\mbox{published\cite{upsohl}} production  cross sections
 for  $\Upsilon(1S)$, $\Upsilon(2S)$, and $\Upsilon(3S)$ states based on a 
data
sample of 16 pb$^{-1}$.  All three states combined yield
 a total of about 1800
candidates reconstructed in the $\mu^+\mu^-$ decay channel.
The D0 collaboration has measured combined $\Upsilon$  cross section 
based on  events 
reconstructed in a 6.6 pb$^{-1}$ data \mbox{sample.\cite{upsd0}} 

\begin{figure} 
\center
\psfig{figure=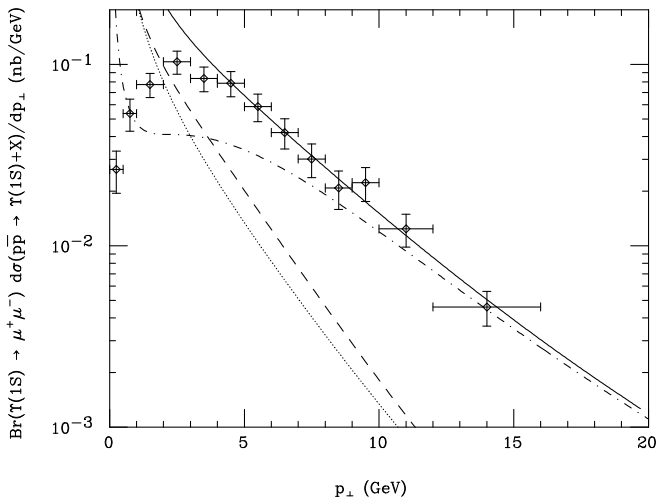,height=2.5in} 
\caption{$\Upsilon(1S)$
differential cross section, compared to the theoretical
prediction.  The dotted line shows the color singlet contribution,
while the dashed lines are the color octet components fitted to the data.
The solid line is the sum of all contributions.
\label{myfig7}}
\end{figure}

     Recently, a theoretical prediction including color-octet contributions was
fitted to the $\Upsilon(1S)$ and $\Upsilon(2S)$ differential 
\mbox{distributions.\cite{cho}}  Figure \ref{myfig7} shows $\Upsilon(1S)$
fit results, which describe the shape of the
$p_T$ distribution well.

     Increased statistics using CDF's full 110 pb$^{-1}$ data sample will allow
finer binning in $p_T$, improving the experimental description of the shape. 
Reconstruction of the $\chi_b$ states via $\Upsilon\gamma$ decay --- while
difficult due to the small mass difference between $\chi_b$ and $\Upsilon$ ---
would provide an additional probe of the underlying production mechanisms.

\section{Conclusion}

Measurements of the differential
 production cross sections for $J/\psi$,
$\psi(2S)$, and three $\Upsilon$ states
have been made at the Tevatron Collider. The prompt and $b$-decay components of
both charmonium states have been extracted.  The prompt $J/\psi$ cross section
has been further subdivided into its direct and $\chi_c$ components.

     These measurements provided the impetus for new theoretical models,
such as the color-octet model, which show potential to explain charmonium
production in $p\overline{p}$ collisions.  Additional experimental results,
such as measurement of the $\psi(2S)$ polarization and reconstruction of
$\chi_b$ states may provide
additional insight into the underlying production mechanisms.

\end{document}